\documentclass[aps,pre,reprint,superscriptaddress,showpacs]{revtex4-1}
\usepackage{mathtools}
\usepackage{bm}
\usepackage{amsmath}

\begin{document}

\title{The structure factor of fat deterministic Vicsek fractals: a small-angle scattering study}

\author{E. M. Anitas}
\email[e-mail:~]{anitas@theor.jinr.ru}
\affiliation{Joint Institute for Nuclear Research, Dubna 141980, Russian Federation}
\affiliation{Horia Hulubei National Institute of Physics and Nuclear Engineering, RO-077125 Bucharest-Magurele, Romania}

\date{\today}

\begin{abstract}
We study here the small-angle scattering structure factor for deterministic fat fractals in the reciprocal space. It is shown that fat fractals are exact self-similar in the range of iterations having the same values of the scaling factor, and therefore in each of these ranges all the properties of regular fractals can be inferred to fat fractals. In order to illustrate the above findings we introduce deterministic "fattened" versions of Vicsek deterministic fractals. We calculate the mono- and polydisperse structure factor and study its scattering properties.
\end{abstract}

\maketitle

\section{Introduction}
Small-angle scattering (SAS; X-ray, neutron, light)~\cite{glatter82:book}
is one of the most used (and quite often, the only possible) method for direct determination of structural properties between $1~\mathrm{nm}$ and $1000~\mathrm{nm}$. 
Since in many investigated micro-objects the presence of certain types of self-similar patterns has been clearly established, such as in various biological and magnetic structures~\cite{gebhardtJACR14,anitasJAC14,naitoEPJB14},  semiconductors~\cite{choJIEC14} or elastomeric membranes~\cite{anitas09}, the natural framework able to describe the scaling behavior is fractal geometry~\cite{mandelbrot83:book,gouyet96:book}. 

Using deterministic fractal models in performing analytical treatment of experimental SAS data with a \textit{single} power-law decay allows us to obtain additional information (fractal form/structure factor, radius of gyration, number of particles inside the fractal, scaling factors)~\cite{chernyJSI10,chernyPRE11,chernyJACR14}. However, numerous examples of experimental SAS data are characterized by a \textit{succession} of power-law decays with decreasing values of scattering exponents~\cite{zhao09,headen09,golosova12}, and it was not until recently when it was shown that such a succession can be described in terms of fat fractal structures (fractals with positive Lebesgue measure)~\cite{anitasEPJB14}. 

Therefore, the aim of this paper is two-fold: first, to present the underlying mechanisms leading to these behaviors (single power-law vs. a succession of power-laws), and second to investigate under which conditions the self-similarity property of regular fractals can be inferred to fat fractals.

\section{Construction of regular and fat fractals}
A detailed description for the construction of regular and fat Cantor fractals, as well as for regular Vicsek fractals has been given in~\cite{chernyJACR10,anitasEPJB14,chernyPRE11}. In this section we summarize their main properties, present the construction of a fattened version of Vicsek fractals and emphasize the differences between regular and fat fractals. In constructing a regular fractal we proceed as follows: we start with an initial cube of edge $l_{0}$, called initiator ($m=0$, $m$ being the fractal iteration number) which is divided into 27 smaller cubes with side length $\beta_{\mathrm{s}} l_{0}$, with the scaling factor given by $\beta_{\mathrm{s}} = (1-\gamma_0)/2$, where $\gamma_{0}$ is the fraction of the removed length.

\begin{figure}
\begin{center}
\includegraphics[width=\columnwidth]{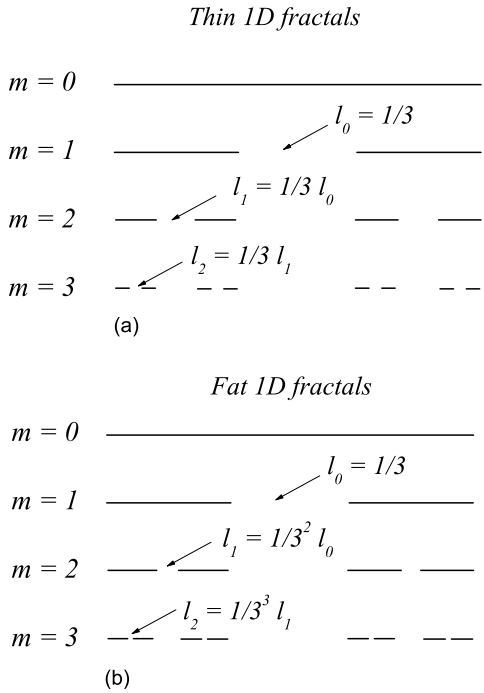}
\end{center}
\caption{Initiator and first three iterations of a generic $1D$ fractal with $\beta_{\mathrm{s}}=1/3$ ($\gamma_{0} = 1/3$): a) thin; b) fat.}
\label{fig:fig1}
\end{figure}

The fattened version of these regular fractals is obtained by considering that for $m=1,2,3$ we remove the fraction $\gamma_{0}$, for iterations $m=2,3,4$ we remove the fraction $\gamma_{1}$ and so on, with
$\gamma_{m}= \alpha^{p_{m}}$, $1/3<\alpha<1$, and the exponent $p_{m}$ is given by the formula $p_1 = p_2 = p_3 = 1,~\cdots,~p_{3k-2} = p_{3k-1} = p_{3k} = k$. The resulting fractal is topologically equivalent to the regular version, but the holes decrease in size sufficiently fast so that, when $m \rightarrow \infty$, the fractal has nonzero and finite volume, and fractal dimension 3. Although the number of cubes at $m$-th iteration of the fat fractal is the same as in the case of regular fractal, in the former case the side length of each cube is 
$l_{m}=l_{0}/2^{m}\prod_{i=1}^{m}(1-\gamma_{i})$ and the scaling factor is given by $\beta_{\mathrm{s}}^{(m)}=(1-\gamma_{m})/2$, and whenever the quantity $(...)$ appears in the exponent, it is to be interpreted as an index and not as a power. Fig.~(\ref{fig:fig1}) shows the construction process of a generic $1D$ deterministic regular and fat fractal based on the Cantor set, for the first four iterations. 
Since the scaling factor depends on the iteration, each scale will be characterized by a different fractal dimension, given by $D = \lim_{m \rightarrow \infty} (\ln N_{m}/\ln \beta_{\mathrm{s}}^{(m)})$, and at each of this scale we shall have a power-law decay $S(q)~\propto~q^{-D}$~\cite{chernyJACR10,chernyPRE11}.

\section{Mono- and polydisperse structure factor}

Since the deterministic fractal at the $m$-th iteration is composed of $N_{m}$ identical units with form factor $F_{0}(\bm{q})$, the form factor of the fractal can be written as~\cite{chernyPRE11} $F(\bm{q})=\rho_{\bm{q}}F_{0}(q a)/N_{m}$ where $a$ is a measure of the size of the fractal's units, $\rho_{\bm{q}}=\sum_{j}e^{-i\bm{q}\bm{r}_{j}}$ is the Fourier component of the density of the composing units of the fractal and $r_{j}$ are the center-of-mass positions of fractal's units. Then, the intensity can be written as~\cite{chernyPRE11} $I(q)=I(0)S(q)/N_{m}\left| F_{0}(q a) \right|^{2}$ where $I(0) = n \left| \Delta \rho \right|^{2} V^{2}$. By definition the fractal structure factor is given by $S(q) \equiv \left\langle \rho_{\bm{q}}\rho_{-\bm{q}} \right\rangle /N_{m}$, and taking into account that the size of the fractal scattering unit is smaller than the minimal distance ($l_{\mathrm{min}}$) between the scattering units, then in the fractal region ($1/l_{0} \lesssim q \lesssim 1/l_{\mathrm{min}}$) we have $F_{0}(q a) \simeq 1$. The analytical expression for the fractal structure factor can be obtained by using the generative function~\cite{chernyJACR10,chernyPRE11,anitasEPJB14}, which gives the positions of the units inside the fractal, and it can be written explicitly as
\begin{equation}
G_{i}(\bm{q}) = \frac{1}{9}\left( 1 + \cos(q_{x}u_{i})\cos(q_{y}u_{i})\cos(q_{z}u_{i}) \right),
\label{eq:gf}
\end{equation}
where $u_{i}=\frac{1}{2}l_{0} (1-\beta_{\mathrm{s}}^{(i)})\prod_{j=1}^{i-1}\beta_{\mathrm{s}}^{(j)}$ and $G_{0}(\bm{q}) \equiv 1$. 

Since it is known that the fat fractal form factor at the $m$-th iteration can be written as $F_{m}(\bm{q})=F_{0}(q\prod_{i=1}^{m}\beta_{\mathrm{s}}^{(i)})\prod_{i=1}^{m}G_{i}(qu_{i})$ ~\cite{anitasEPJB14}, we can use this result to write an explicit expression for the Fourier component of the density of fractal units $\rho_{\bm{q}}=N_{m}\prod_{i=1}^{m}G(\bm{q}u_{i})$. Finally, we can write the fat fractal structure factor in the following form $S(q)=N_{m} \left \langle \prod_{i=1}^{m} \left| G_{i}(\bm{q}u_{i})\right|^{2} \right \rangle$. 

\begin{figure}
\begin{center}
\includegraphics[width=\columnwidth]{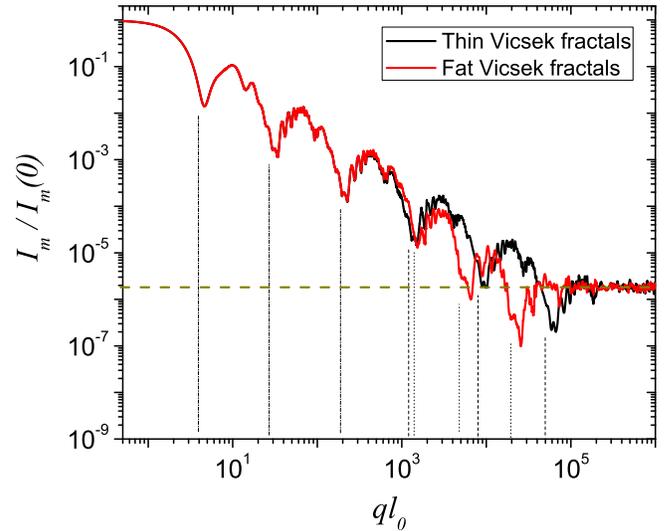}
\end{center}
\caption{ (Color online)
Monodisperse fractal structure factor for the sixth iteration and $\gamma_{0} = 0.7$. Black (dark): regular fractal; Red (gray): fat fractal; Horizontal dashed line: asymptotic values; Green (light gray) vertical lines: common positions of minima for both, regular and fat fractals; Black (dark) vertical dashed lines: positions of minima of regular fractals; Red (gray) vertical dotted lines: positions of minima of fat fractals.}
\label{fig:fig2}
\end{figure}

For polydispersity, we consider here a log-normal distribution function $D_{\mathrm{N}}(l)$ which gives the probability of finding a fractal whose size is in the interval $(l,l+dl)$~\cite{chernyPRE11}.  Therefore the polydisperse structure factor can be computed using (Fig.~\ref{fig:fig3}) $S(q)=N_{m}\int_{0}^{\infty} \left \langle \prod_{i=1}^{m} \left| G_{i}(\bm{q}u_{i})\right|^{2} \right\rangle D_{N}(l)dl$.

\section{Results and discussions}
\begin{figure}
\begin{center}
\includegraphics[width=\columnwidth]{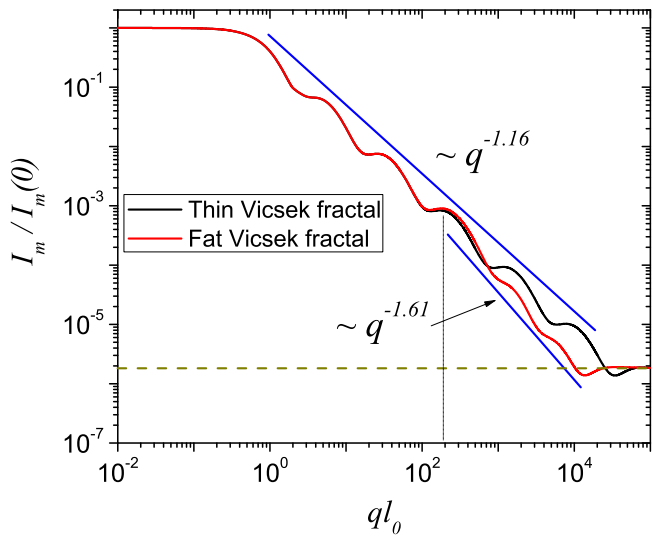}
\end{center}
\caption{(Color online) Polydisperse fractal structure factor for the sixth iteration, $\gamma_{0}=0.7$ and relative variance $\sigma_{\mathrm{r}}=0.4$. Black (dark): regular fractal; Red (gray): fat fractal; Horizontal dashed line: asymptotic values; Black (gray) dotted vertical line indicates the point where takes place the changes in the slopes of power-law decay.}
\label{fig:fig3}
\end{figure}

The numerical results at $m=6$ for both, monodisperse and polydisperse fractal structure factor, at fixed $\gamma_{0}$ are shown in Fig.~\ref{fig:fig2} and respectively, in Fig.~\ref{fig:fig3}.

We can observe that the monodisperse structure factor is characterized by three main regions on a double-logarithmic scale: a plateau in the range $ql_{0} \lesssim 1$ (Guinier region), \textit{a succession} of generalized power-law decays at $1 \lesssim ql_{0} \lesssim l_{0}/u_{m}$ (intermediate region) and an asymptotic region at $1/u_{m} \lesssim q$. The position of minima in the intermediate region are obtained when the fractal units interfere out of phase, and since the most common distances between the center of mass of the units are given by $2u_{m}$, imposing the condition $2u_{m}=\pi/q$, allows us to write the position of minima of the fat fractal structure factor (Green; light gray and Red; dark gray) vertical lines in Fig.~\ref{fig:fig2} in the form $q_{k}l_{0}\simeq\pi/((1-\beta_{\mathrm{s}}^{(k)})\prod_{i=1}^{k}{\beta_{\mathrm{s}}^{(i)}})$. When all the scaling factors $\beta_{\mathrm{s}}^{(i)}$ are equal, it gives the minima positions of the corresponding regular fractal (Black (dark) vertical lines in Fig.~\ref{fig:fig2} (see~\cite{chernyJACR10,chernyPRE11}).

When the polydispersity is taken into account in calculating the fractal structure factor, the minima and maxima in the intermediate regions became smeared out (Fig.~\ref{fig:fig3}) as expected and therefore the fat fractal structure factor (Red; dark gray) curves in Fig.~\ref{fig:fig3} is characterized now by \textit{a succession} of simple power-law decays with decreasing values of the scattering exponents, as seen in many SAS experimental data. Note that a succession of SAS data with increasing values of scattering exponents can be described using multi-phase structures~\cite{chernyJACR14}.

In the asymptotic region the monodisperse and polydisperse fat fractal structure factor (Red; dark gray) curve) is $S(q) \simeq 1$, and therefore the asymptotic values will tend to $1/N_{m}$, as in the case of regular fractals (Black (dark) curves)~\cite{chernyJACR10,chernyPRE11}.

\section{Conclusions}
We have shown that a succession of power-law decays (either simple or generalized) of small-angle scattering data from fat fractals, arise as a result of a repeated increase of the scaling factor, after a finite number of iterations. In addition, for the iterations with identical values of the scaling factor we have shown that fat fractals resemble exactly the behavior of regular fractals, and therefore the former can be considered as self-similar for these iterations. The above findings have been illustrated by calculating the mono- and polydisperse scattering structure factors for a newly developed deterministic Vicsek fat fractal.

\end{document}